\documentclass[twocolumn,showpacs,preprintnumbers,amsmath,amssymb]{revtex4}%
\usepackage{graphicx}
\usepackage{dcolumn}
\usepackage{bm}
\usepackage{amsmath}
\usepackage{amsfonts}
\usepackage{amssymb}%
\begin{document}
\title{Quantum computing with collective ensembles of multi-level systems}
\author{E. Brion}
\affiliation{Lundbeck Foundation Theoretical Center for Quantum System Research, Department
of Physics and Astronomy, University of Aarhus, DK-8000 \AA rhus C, Denmark.}
\author{K. M\o lmer}
\affiliation{Lundbeck Foundation Theoretical Center for Quantum System Research, Department
of Physics and Astronomy, University of Aarhus, DK-8000 \AA rhus C, Denmark.}
\email{moelmer@phys.au.dk}
\author{M. Saffman}
\affiliation{Department of Physics, University of Wisconsin, 1150 University Avenue,
Madison, Wisconsin 53706, USA.}
\date{\today}

\begin{abstract}
We propose a new physical approach for encoding and processing of quantum information in
ensembles of multi-level quantum systems, where the different bits are
not carried by individual particles but associated with the collective
population of different internal levels. One- and two-bit gates are
implemented by collective internal state transitions taking place in
the presence of an excitation blockade mechanism which restricts the
population of each internal state to the values zero and unity. 10-20 bit quantum computers can be built via this scheme in single trapped clouds of ground state atoms subject to the Rydberg excitation blockade mechanism, and the linear dependence
between register size and the number of internal quantum states in
atoms offers realistic means to reach larger registers.

\end{abstract}
\maketitle

\begin{figure}
[ptb]
\begin{center}
\includegraphics[
height=2.0003in,
width=3in
]%
{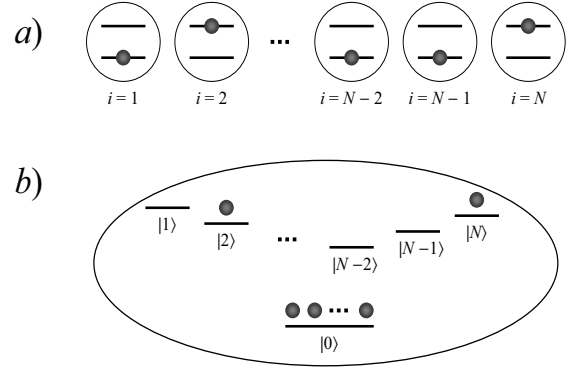}%
\caption{Encoding of qubits in two-level and multi-level atoms. a)
Conventional qubit encoding of the $N$-bit state $\left\vert 01\ldots
001\right\rangle $ in $N$ two-level systems. b) Qubit encoding in the
symmetric states of an ensemble of $(N+1)$--level systems. The state
$\left\vert 0\right\rangle $ is a reservoir state, populated initially by all
atoms. Bullets represent the number of atoms populating the different single
particle states to encode the $N$-bit state $\left\vert 01\ldots
001\right\rangle $.}%
\label{Fig1}%
\end{center}
\end{figure}

Quantum computers have the potential to solve a number of difficult
mathematical problems such as factoring and unstructured search much
more efficiently than their classical counterparts \cite{1,2,3}.
This power stems mainly  from the quantum superposition principle,
which allows a single system to simultaneously explore the entire
(computational) state space. Although impressive quantum state
control has been demonstrated in very high dimensional systems such
as Rydberg state manifolds \cite{4,5} and molecules \cite{6,7}, the
number of qubits one can store in such individual systems scales
only logarithmically with the Hilbert space dimension: for instance, a
thousand states correspond to the Hilbert space dimension of only
ten logical bits $2^{10}=1024$. Genuine scalability aiming at
several tens of logical bits can thus only be reached through
composing individual quantum systems. Most quantum
computing proposals use a large number of two-level quantum
systems, each representing a physical qubit. The tensor product
space of $N$ such particles accommodates $N$ qubits of information
(Fig. \ref{Fig1}a), and a quantum algorithm is carried out with
sequences of unitary operations on single qubits and pairs of
qubits, the so-called universal quantum gates \cite{8}.

Controlling the joint state of many particles with suitable
interactions to mediate two-bit gates between any pair of qubits is
an outstanding challenge. In this Letter, we propose a new encoding
and computing strategy that uses an ensemble of collectively
addressed identical quantum systems each of which has $(N+1)$ long
lived internal states $\left\vert i\right\rangle $, $i=0,\ldots,N$.
We shall regard the $\left\vert i=0\right\rangle $ state as a
reservoir state, initially populated by all members of the ensemble,
and formally associate the computational register state $\left\vert
b_{1},b_{2},\ldots,b_{N}\right\rangle $ $\left(  b_{i}=0,1\right) $\
with the symmetric state of the ensemble with $b_{i}$ ensemble
members populating state $\left\vert i\right\rangle $ (Fig.
\ref{Fig1}b). There is thus a direct correspondence between the
binary representation of a register state and the Fock state
representation, providing the zero and unity populations of different
single particle states, so that, in principle, only $K=N$ ensemble
members suffice to represent all $2N$ possible register states. In
practice, however, ensemble sizes about an order of magnitude larger
than the register size, or even more, improve a number of properties
of our proposal, and do not set major experimental problems:
ensembles of thousands of atoms are routinely produced and
manipulated in quantum optics laboratories.
%TCIMACRO{\FRAME{ftbpFU}{3in}{2.0003in}{0pt}{\Qcb{Encoding of qubits in
%two-level and multi-level atoms. a) Conventional qubit encoding of the $N$-bit
%state $\left\vert 01\ldots001\right\rangle $ in $N$ two-level systems. b)
%Qubit encoding in the symmetric states of an ensemble of $(N+1)$--level
%systems. The state $\left\vert 0\right\rangle $ is a reservoir state,
%populated initially by all atoms. Bullets represent the number of atoms
%populating the different single particle states to encode the $N$-bit state
%$\left\vert 01\ldots001\right\rangle $.}}{\Qlb{Fig1}}{fig1.eps}%
%maintain-aspect-ratio TRUE;  display "USEDEF";  valid_file "F";  width 3in;
%height 2.0003in;  depth 0pt;  original-width 0pt;  original-height 0pt;
%cropleft "0";  croptop "1";  cropright "1";  cropbottom "0";
%filename 'Fig1.eps';file-properties "XNPEU";}}}%
%BeginExpansion
%EndExpansion

How can we prepare symmetric ensemble states and perform logical
operations with at most one particle in each long-lived register
state? In the case of atomic ensembles, our candidate solution to
this problem is the so-called Rydberg blockade \cite{9}. This
mechanism is a consequence of the large dipole-dipole interaction
\cite{10} which strongly shifts the energies of Rydberg excited
atoms with separations of several $\mu m$, so that the presence of
one Rydberg atom is enough to prevent the excitation of all other
atoms in the ensemble \cite{11}. This blockade is able to entangle
two atoms and can thus be used to achieve two-bit quantum gates in
individual-atom systems \cite{9}. It has also been suggested as a
means to encode one qubit of information in a mesoscopic ensemble of
two-level atoms \cite{12} and perform arbitrary rotations of the
qubit. Finally, it allows the implementation of a conditional quantum
gate on the qubits stored in two different ensembles, either
directly if the ensembles are within the range of the dipole-dipole
interaction, or by transferring the states of both ensembles into a
single intermediate ensemble and carrying out the gate here, before
transferring the (entangled) qubits back to their original ensembles
\cite{12}. The coupling of the mesoscopically populated reservoir
state to the symmetric state with precisely a single atom
transferred to another state is enhanced by the factor $\sqrt{K}$
compared to the single atom coupling ($K$ is the number of atoms in
the ensemble), and may hence be strong enough to yield appreciable
coupling even to field states with only a single photon: transfer of
quantum states between samples may thus be achieved by exchange of
photons. Although it also relies on Rydberg blockade, our scheme
differs fundamentally by encoding the entire $N$-bit quantum
register in a single collectively addressed mesoscopic ensemble. 

\begin{figure}
[ptb]
\begin{center}
\includegraphics[
height=1.5in,
width=2.5in
]%
{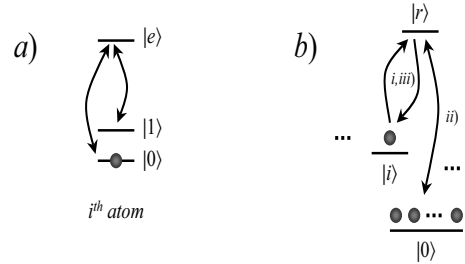}%
\caption{One-bit gates in two-level and multi-level atoms. a) Qubit rotation
of the $i$th qubit through a Raman process via an excited state of the $i$th
atom. b) Qubit rotation of the $i$th qubit in the ensemble representation.
$i$) a $\pi$-pulse transfers the level population in $\left\vert
i\right\rangle $ to $\left\vert r\right\rangle $, $ii$) coherent coupling in
the two-state system with zero and one atom in $\left\vert r\right\rangle $,
$iii$) a $\pi$-pulse transfers the level population in $\left\vert
r\right\rangle $ to $\left\vert i\right\rangle $.}%
\label{Fig2}%
\end{center}
\end{figure}

Let us now show how to implement one- and two-bit gates in an ensemble of
$(N+1)$-level atoms, making use of optical transitions and the Rydberg
blockade mechanism. In mesoscopic ensembles, single-qubit rotations are a
little more complex in their structure than in a single-atom implementation, as illustrated in
Fig. \ref{Fig2}. Rather than coupling qubit levels $\left\vert 0\right\rangle
$ and $\left\vert 1\right\rangle $ by a site specific coherent, e.g., Raman,
process (Fig. \ref{Fig2}a), a coherent coupling of states with zero and one
Rydberg excited atoms enables rotations in that space, and swapping the
population between any atomic level $\left\vert i\right\rangle $ and the
Rydberg state before and after this unitary rotation, effectively implements
the rotation on the $i$th qubit (Fig. \ref{Fig2}b).%

%TCIMACRO{\FRAME{ftbpFU}{3in}{2.0003in}{0pt}{\Qcb{One-bit gates in two-level
%and multi-level atoms. a) Qubit rotation of the $i$th qubit through a Raman
%process via an excited state of the $i$th atom. b) Qubit rotation of the $i$th
%qubit in the ensemble representation. $i$) a $\pi$-pulse transfers the level
%population in $\left\vert i\right\rangle $ to $\left\vert r\right\rangle $,
%$ii$) coherent coupling in the two-state system with zero and one atom in
%$\left\vert r\right\rangle $, $iii$) a $\pi$-pulse transfers the level
%population in $\left\vert r\right\rangle $ to $\left\vert i\right\rangle $.}%
%}{\Qlb{Fig2}}{fig2.eps}{\special{ language "Scientific Word";
%type "GRAPHIC";  maintain-aspect-ratio TRUE;  display "USEDEF";
%valid_file "F";  width 3in;  height 2.0003in;  depth 0pt;
%original-width 0pt;  original-height 0pt;  cropleft "0";  croptop "1";
%cropright "1";  cropbottom "0";  filename 'Fig2.eps';file-properties "XNPEU";}%
%}}%
%BeginExpansion
%EndExpansion

Here, we assume that any of the stable single atom states can be coupled
coherently to the Rydberg state
$\left\vert r\right\rangle $, or to different Rydberg states $\left\vert r\right\rangle $,
$\left\vert r^{\prime}\right\rangle $... via suitable intermediate excited
states. This is not a trivial task, since optical fields in general can
resonantly excite all Zeeman sub-states in the ground state manifold, and if
such states are used to encode the qubit register, we may need to apply
external fields to lift their degeneracy or to apply techniques from control
theory and act with shaped pulses which leave all register states untouched,
and only excite the relevant one. A possible implementation using a
homogeneous magnetic field to Zeeman select the desired states is described below.

At this point it is important to recall that the coherent driving of
the transition to the $i$th register level is enhanced by the
symmetric coupling to all the single particle components of the
collective state, but unlike previous analyses \cite{12}, this
enhancement is not the square root of the number of atoms
$\sqrt{K}$, which is a fixed number, but only of the number of atoms
available for the transition, $\sqrt{K-\sum_{j\neq i}b_{j}}$,
because the population residing in the other register states is not
coupled. During quantum computation, the ensemble will populate a
superposition of states with accordingly different values of the
coupling strengths and thereby introduce an inhomogeneity in the
system. If $N\ll K$, variations in the coupling parameters may be
significantly reduced and, if needed, simple composite pulses
\cite{13} can ensure robustness against these small variations.

\begin{figure}
[ptb]
\begin{center}
\includegraphics[
height=1.5in,
width=2.5in
]%
{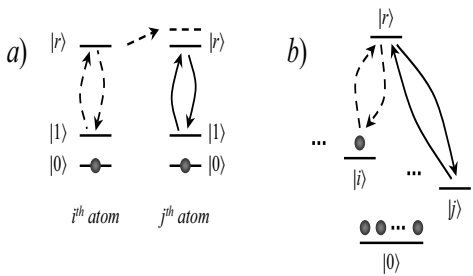}%
\caption{Two-bit gates in two-level and multi-level atoms. a) Two-bit gate via
Rydberg blockade. If the $i$th qubit level $\left\vert 1\right\rangle _{i}$ is
excited to $\left\vert r\right\rangle _{i}$, the transition $\left\vert
1\right\rangle _{j}\rightarrow\left\vert r\right\rangle _{j}$ is shifted out
of resonance, and a $2\pi$-pulse on the $j$th atom yields a conditional phase.
b) Two-qubit gate in the ensemble representation. After a $\pi$-pulse on the
transition $\left\vert i\right\rangle \rightarrow\left\vert r\right\rangle $,
the transition $\left\vert j\right\rangle \rightarrow\left\vert r\right\rangle
$ is blocked conditioned on the atomic population in $\left\vert
i\right\rangle $, and a $2\pi$-pulse yields a conditional phase.}%
\label{Fig3}%
\end{center}
\end{figure}

As sketched in Fig. \ref{Fig3}, the implementation of two-bit gates in our
ensemble scheme is very different from the single-atom Rydberg blockade
proposal, where the dipole-dipole interaction disturbs the resonance condition
when two different atoms are exposed to resonant driving fields (Fig.
\ref{Fig3}a). In our scheme, the excitation of the Rydberg state from a
logical \textquotedblleft$1$\textquotedblright\ of the $i$th qubit,
\textit{i.e.}, from a single atom in the state $\left\vert i\right\rangle $,
prevents the resonant driving of an atom in register state $\left\vert
j\right\rangle $, and hence a conditional phase or NOT gate can be applied to
the $j$th qubit (Fig. \ref{Fig3}b).%

%TCIMACRO{\FRAME{ftbpFU}{3in}{2.0003in}{0pt}{\Qcb{Two-bit gates in two-level
%and multi-level atoms. a) Two-bit gate via Rydberg blockade. If the $i$th
%qubit level $\left\vert 1\right\rangle _{i}$ is excited to $\left\vert
%r\right\rangle _{i}$, the transition $\left\vert 1\right\rangle _{j}%
%\rightarrow\left\vert r\right\rangle _{j}$ is shifted out of resonance, and a
%$2\pi$-pulse on the $j$th atom yields a conditional phase. b) Two-qubit gate
%in the ensemble representation. After a $\pi$-pulse on the transition
%$\left\vert i\right\rangle \rightarrow\left\vert r\right\rangle $, the
%transition $\left\vert j\right\rangle \rightarrow\left\vert r\right\rangle $
%is blocked conditioned on the atomic population in $\left\vert i\right\rangle
%$, and a $2\pi$-pulse yields a conditional phase.}}{\Qlb{Fig3}}{fig3.eps}%
%maintain-aspect-ratio TRUE;  display "USEDEF";  valid_file "F";  width 3in;
%height 2.0003in;  depth 0pt;  original-width 0pt;  original-height 0pt;
%cropleft "0";  croptop "1";  cropright "1";  cropbottom "0";
%filename 'Fig3.eps';file-properties "XNPEU";}}}%
%BeginExpansion
%EndExpansion

\begin{figure}
[ptb]
\begin{center}
\includegraphics[
width=3.3in
]%
{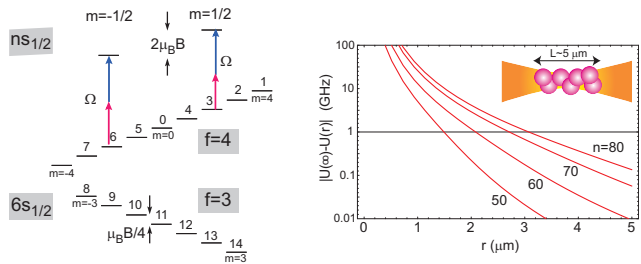}%
\caption{(Color online) Cesium level scheme and identification of qubit register. Encoding of
reservoir state $0$ and $14$ register states in the Zeeman ground states of
Cs. Coupling of $\left\vert 3\right\rangle =\left\vert f=4,m=2\right\rangle $
and $\left\vert 6\right\rangle =\left\vert f=4,m=-2\right\rangle $ to Rydberg
states is shown. The plot shows the dipole-dipole interaction strength versus
atom separation $r$ for Rydberg states with $n=50-80$. As indicated in the
inset an atomic ensemble can be confined in an optical trap with a length
scale of $5~\mu \rm m$.}%
\label{Fig4}%
\end{center}
\end{figure}

Fig. \ref{Fig4} shows specific implementation of the above ideas for Cs atoms, whose nuclear
spin $I=7/2$ provides $16$ stable Zeeman states in the $f=3,4$ hyperfine levels. For effective Rydberg blockade the atoms are
transferred by a two-photon excitation to a high lying $ns$ state with
$n \sim 70$. The hyperfine structure of the Rydberg level is unresolved
so that the ground states are coupled to fine structure states $\left\vert
ns_{1/2},m_{j}=\pm1/2\right\rangle $. Application of a magnetic field $B$ to the
atomic sample shifts all transition frequencies so that the only degenerate
transitions are $\left\vert 6s_{1/2},f=4,m=-4\right\rangle \leftrightarrow
\left\vert ns_{1/2},m_j=-1/2\right\rangle $ and $\left\vert 6s_{1/2}%
,f=4,m=4\right\rangle \leftrightarrow\left\vert ns_{1/2},m_j=1/2\right\rangle $.
We exclude one of these states,
leaving $15$ ground states available for the reservoir and encoding of up to
$14$ qubits. Each of the ground to Rydberg transitions are separated by at
least $\frac{\mu_{B}B}{4\hbar}$, with $\mu_{B}$ the Bohr magneton, and as long
as this quantity is large compared to the two-photon excitation frequency
$\Omega$ any ground state can be selectively excited by appropriate choice of
the laser frequencies. Typical numbers set by the need to respect the finite
lifetime of the Rydberg level are \cite{14}$\ \Omega/2\pi \sim 1~\rm MHz$ so
that a modest field of $B \sim 15~\rm G$ will give suppression of undesired
excitation at the $1\%$ level.%

%TCIMACRO{\FRAME{ftbpFU}{3in}{2.0003in}{0pt}{\Qcb{Cesium level scheme and
%identification of qubit register. Encoding of reservoir state $0$ and $14$
%register states in the Zeeman ground states of Cs. Coupling of $\left\vert
%3\right\rangle =\left\vert f=4,m=2\right\rangle $ and $\left\vert
%6\right\rangle =\left\vert f=4,m=-2\right\rangle $ to Rydberg states is shown.
%The plot shows the dipole-dipole interaction strength versus atom separation
%$r$ for Rydberg states with $n=50-80$. As indicated in the inset an atomic
%ensemble can be confined in an optical trap with a length scale of $5\mu m$.}%
%}{\Qlb{Fig4}}{fig4.eps}{\special{ language "Scientific Word";
%type "GRAPHIC";  maintain-aspect-ratio TRUE;  display "USEDEF";
%valid_file "F";  width 3in;  height 2.0003in;  depth 0pt;
%original-width 0pt;  original-height 0pt;  cropleft "0";  croptop "1";
%cropright "1";  cropbottom "0";  filename 'fig4.eps';file-properties "XNPEU";}%
%}}%
%BeginExpansion
%EndExpansion

In an ensemble with random atomic positions blockade is required 
between pairs of atoms with either the same or opposite signs of
$m$, i.e. $m_j=\{\pm 1/2,\pm 1/2\}$ or $m_j=\{\pm 1/2, \mp 1/2\}$.  In both cases an effective long range
interaction  occurs via the F\"{o}rster
mechanism \cite{12} due to the near resonance of two-atom states
$ns+ns\leftrightarrow np_{1/2(3/2)}+(n-1)p_{1/2(3/2)}$. Although the
energy defects vary by about a factor of three at large $n$ for different combinations 
of the $p_{1/2}$ and $p_{3/2}$ states, it can be shown that the interaction is 
always strong and deviates from  isotropic by only about 15\% 
for atomic separations greater than
several $\mu\rm m$ with the approximate form \cite{15}
$U=\pm\sqrt{\left( \frac{\hbar\delta}{2}\right)  ^{2}+\frac
{4}{3}\frac{C_{3}^{2}}{r^{6}}}$ with $\delta$ an effective average 
two-atom energy
defect and $C_{3}=\frac{e^{2}}{4\pi\epsilon_{0}}\left\langle
np\left\Vert r\right\Vert ns\right\rangle \left\langle \left(
n-1\right)  p\left\Vert r\right\Vert ns\right\rangle $. We see from
Fig. \ref{Fig4} that by excitation of $n=70$ we obtain close to $1~\rm GHz$
of interaction at $3~\mu\rm m$ and about $80~\rm MHz$ at separations as large 
as $5~\mu\rm m$.
This is sufficient for high fidelity blockade of small samples of up
to several hundred atoms. It may be advantageous to apply an optical
lattice potential with only a single atom per site. The atoms would
still not require individual addressing but the trapping may reduce
the effect of collisions among the atoms, while many atoms will be
within the range of the dipole-dipole interaction.

Having both one- and two-bit gates at our disposal, and the ability to
initialize the system, we only need to specify an effective read-out
mechanism. This can for example be achieved by use of the Rydberg blockade
\cite{16}, or by coupling the register levels in a controlled manner to
excited states from which, \textit{e.g.}, ionization can be observed.

An implementation of our proposal with, say, $50-100$ cold trapped
cesium atoms in a far-off resonant optical trap may allow quantum
computing with $14$ qubits. This would exceed the performance of any
other proposal implemented to date. It is necessary, however, to
develop strategies to scale the model to even larger numbers of
qubits. To this end, we emphasize the linearity reported between single-particle
internal Hilbert space dimension and number of qubits encoded. It is
conceivable that atoms, or molecules, with more stable states can be
used and approach the $100$-bit regime, but the perhaps simplest
approach may be to place a small number, say $10-20$, of such
samples so that they can be addressed by different laser fields, but
still affect each other by the Rydberg blockade mechanism. Also,
superconducting transmission lines work well with the large dipole
moments of Rydberg excited states \cite{18}, and the use of
mesoscopic ensembles of atoms also provides an efficient coupling to
single photons \cite{19,20}: both types of coupling may be used to
communicate between different ensembles constituting a larger
quantum computer. The number of qubits encoded simply add, and the
set of one- and two-bit gates is equally straightforward to
implement. Another, perhaps intriguing, approach towards larger
registers is to consider mixtures of different species, which may be
contained within the same volume, and where the Rydberg blockade may
also apply between species. In this case, the total number of qubits
is also found by adding the contributions from each species.

Though we discussed here an implementation of our encoding scheme making use of
the Rydberg blockade mechanism to prevent transitions of an atomic
ensemble out of the register Hilbert space, other physical systems can be considered which
exhibit excitation blockade: for instance, hybrid schemes for quantum
computing where an ensemble of atoms or molecules interacts with a
single saturable two-level system (e.g., a Cooper pair box
\cite{22}) are promising candidates for implementation of our
encoding scheme. The challenge is now to find ways to reliably
control the internal state transitions and to identify the most
efficient one- and two-bit gates. Further natural developments
involve the analysis of the most relevant errors and the
achievements of appropriate error-correcting codes.

In conclusion, we have proposed an approach to quantum computing
that makes effective use of the single particle Hilbert space
dimension by encoding an $N$-bit quantum register in particles with
only $(N+1)$ internal states. The principles are new by the
effective concentration of quantum information in single particle
multi-level systems. We emphasize, however, that, in our approach,
we retain the basic qubit structure, which is essential for easy
implementation of quantum gates. Conventional tensor-product state
encoding requires individual access to the atoms and highly
controlled mutual interactions between them, and a high degree of
entanglement is present during computation. Instead, we use the
collective population of levels, and all operations are carried out
by only collective and symmetric access to the atoms. Interactions
also play an important role and entanglement is present in our
proposal, but in a very different manner. Thus, for example, the
\textquotedblleft classical\textquotedblright\ register states,
which are simple product states, cf. Fig. \ref{Fig1}a are now
entangled states, cf. Fig. \ref{Fig1}b. These physical and
mathematical differences with conventional encoding will have
important consequences for the further development of optimal
algorithms, and \textit{e.g.}, for error-correcting codes
\cite{1,17}, where the typical errors on different qubits will
definitely not be independent - the atoms will not lose a level, but
the ensemble can lose a single atom, which calls for different
correction measures.

This work was supported by ARO-DTO, NSF, and the European Union integrated project SCALA.

\end{document}